\documentstyle[12pt]{article}
\makeatletter
\@addtoreset{equation}{subsection}
\makeatother

\begin{document}
\begin{center}
{\Large \bf
Classical and Quantum Fields in Curved Spacetime:} \\[0.5cm]
{\large\bf Canonical Theory versus Conventional Construction}\\[1.5cm]
{\bf Vladimir S.~MASHKEVICH}\footnote {E-mail:
dima@mashke.org}  \\[1.4cm]
{\it 25 Manor Drive Apt. 5-P\\
Newark NJ 07106\\
USA }\footnote {Present address} \\[0.4cm]
{\it and} \\[0.4cm]
{\it Institute of Physics, National academy of sciences
of Ukraine\\  252028 Kiev, Ukraine}
 \\[1.4cm]
\vskip 1cm

{\large \bf Abstract}
\end{center}

We argue that the conventional construction for quantum fields in curved
spacetime has a grave drawback: It involves an uncountable set of physical
field systems which are nonequivalent with respect to the Bogolubov
transformations, and there is, in general, no canonical way for choosing
a single system. Thus the construction does not result in a definite theory.
The problem of ambiguity pertains equally both to quantum and classical
fields. The canonical theory is advanced, which is based on a canonical,
or natural choice of field modes. The principal characteristic feature
of the theory is relativistic-gravitational nonlocality: The field at a
spacetime point $(s,t)$ depends on the metric at $t$ in the whole 3-space.
The most fundamental and shocking result is the following: In the case of
a free field in curved spacetime, there is no particle creation. Applications
to cosmology and black holes are given. The results for particle energies
qqare in complete agreement with those of general relativity. A model of the
universe is advanced, which is an extension of the Friedmann universe; it
lifts the problem of missing dark matter.

\newpage

\section*{Introduction}

The conventional construction in quantum field theory in curved
spacetime is usually presented in such a form [1,2] that the
fulfillment of the canonical commutation relations is not evident.
That was the reason for a criticism against the construction
in our paper [3]. But as it may be seen from a general approach
[4] and will be shown in this paper, the commutation relations
may be fulfilled. This is the answer to the criticism.

But now we argue that the conventional construction has the
following grave drawback: It involves an uncountable set of
physical field systems which are nonequivalent with respect to
the Bogolubov transformations, and there is, in general, no
canonical way for choosing a single system. Thus the conventional
construction does not result in a definite field theory. This
is the reason for our using the term ``construction'' rather than
``theory''.

The problem of ambiguity is not specifically concerned with
quantum fields: for classical ones the situation is the same.
Thus both classical and quantum fields should be treated on
equal terms.

A field as a dynamical variable (observable) is presented in
the form of the expansion in terms of modes, coefficients being
the canonical dynamical variables. The problem of fixing the
field reduces to that of settling the modes.

In the conventional construction, the basic condition on the
field is that it obey the Klein-Gordon equation. Under this
condition, there exists a vast arbitrariness for the mode
settling. It should be particularly emphasized that the
ambiguity involved has nothing to do both with different
representations of the same field connected by the Bogolubov
transformations and with canonical/unitary transformations
connecting equivalent fields.

In the canonical theory, the basic condition is that the mode
settling be canonical, or natural, i.e., based on the spacetime
structure only. Under this condition, the modes are settled
unambiguously, which results in a definite field theory---the
canonical one.

Dynamics is presented both in Liouville/Schr\"odinger and
Hamilton/Heisenberg pictures. The quantum Hamiltonian
implies the existence of particles, their energies being
time dependent. The most fundamental and shocking result
is the following: In the case of a free quantum field
in curved spacetime, there is no particle creation.

The Klein-Gordon equation is violated in the generic case
of a nonstationary metric. A local change in the metric
results in changing modes and frequencies/energies.
We call this phenomenon relativistic-gravitational
nonlocality. Nonlocality is incompatible with the local
principle of covariance. The canonical theory meets the
geometric principle, which is more general: Spacetime
structure and dynamics should be phrased in geometric form.

Applications to cosmology and black holes are given. The
results for particle energies are in complete agreement with
those of general relativity. A model of the universe is advanced,
which is an extension of the Friedmann universe; the model lifts
the problem of the missing dark matter.

\section{Preliminaries: The algebraic description of a
nonautonomous system}

A classical/quantum physical system is described in terms of
dynamical variables, states, and time evolution, or dynamics.

\subsection{Statics: Dynamical variables, states, and mean
values}

A set ${\cal A}$ of dynamical variables $A$,
\begin{equation}
\{A:A\in {\cal A}\},
\label{1.1.1}
\end{equation}
and a set $\Omega$ of states $\omega$, which are functionals
on ${\cal A}$,
\begin{equation}
\left\{ \omega:\omega\in \Omega \right\},
\label{1.1.2}
\end{equation}
are given,
\begin{equation}
\omega(A)
\label{1.1.3}
\end{equation}
being the mean value of a dynamical variable $A$ in a state
$\omega$.

It is convenient to introduce a quantity $\rho$ which is
equivalent to $\omega$,
\begin{equation}
\rho\leftrightarrow\omega,\quad \omega(A)=\left\langle
\rho,A \right\rangle.
\label{1.1.4}
\end{equation}

\subsection{Dynamics: Two pictures}

For a nonautonomous system, a dynamical variable is a given
function of time,
\begin{equation}
A=A(t)\in{\cal A}.
\label{1.2.1}
\end{equation}
There are two pictures for describing the time dependence of
mean values:
\begin{equation}
[\omega(A(t))]_{t}=\omega_{t}(A(t))=\omega(A_{t}(t));
\label{1.2.2}
\end{equation}
for some time $t_{0}$,
\begin{equation}
\omega_{t_{0}}=\omega,\qquad A_{t_{0}}(t_{0})=A(t_{0}).
\label{1.2.3}
\end{equation}

The dynamical equations are
\begin{equation}
\frac{d\rho_{t}}{dt}=-{\rm i}[H(t),\rho_{t}],
\label{1.2.4}
\end{equation}
\begin{equation}
\frac{dA_{t}(t)}{dt}=\left(\frac{dA(t)}{dt}\right)_{t}+
{\rm i}[H_{t}(t),A_{t}(t)],
\label{1.2.5}
\end{equation}
where $H$ is the Hamiltonian.

\subsection{A classical system}

A phase space $\Gamma$ is given,
\begin{equation}
\left\{ \gamma:\gamma\in \Gamma \right\},\qquad
\gamma=\left\{ (\alpha_{j},\alpha_{j}^{*}):
\alpha_{j}\in C,\:j\in J \right\},
\label{1.3.1}
\end{equation}
$\Gamma$ being a differentiable manifold. A dynamical variable
\begin{equation}
A=A(\gamma).
\label{1.3.2}
\end{equation}
$\rho$ is the distribution function,
\begin{equation}
\rho=\rho(\gamma).
\label{1.3.3}
\end{equation}
The mean value
\begin{equation}
\left\langle \rho,A \right\rangle=\int d\gamma\rho(\gamma)
A(\gamma),\qquad d\gamma=\prod _{j}d\alpha_{j}d\alpha_{j}^{*}.
\label{1.3.4}
\end{equation}
For a pure state,
\begin{equation}
\omega_{\gamma_{0}}\leftrightarrow\rho_{\gamma_{0}}(\gamma)
=\delta(\gamma-\gamma_{0}),\qquad \omega_{\gamma_{0}}(A)=
A(\gamma_{0}).
\label{1.3.5}
\end{equation}
${\rm i}[A,B]$ is the Poisson bracket,
\begin{equation}
[A,B]=\sum_{j}\left( \frac{\partial A}{\partial \alpha_{j}}
\frac{\partial B}{\partial \alpha_{j}^{*}}-
\frac{\partial A}{\partial \alpha_{j}^{*}}
\frac{\partial B}{\partial\alpha_{j} } \right)=
\frac{1}{{\rm i}}\sum_{j}\left( \frac{\partial A}{\partial
p_{j}}\frac{\partial B}{\partial q_{j}}
-\frac{\partial A}{\partial
q_{j}}\frac{\partial B}{\partial p_{j}} \right),
\label{1.3.6}
\end{equation}
where
\begin{equation}
\alpha_{j}=\frac{1}{\sqrt{2}}\left( \sqrt{
\omega_{j}}q_{j}+\frac{{\rm i}}{\sqrt{\omega_{j}}}
p_{j} \right),\qquad \alpha_{j}^{*}=
\frac{1}{\sqrt{2}}\left( \sqrt{\omega_{j}}q_{j}
-\frac{{\rm i}}{\sqrt{\omega_{j}}}p_{j} \right),
\label{1.3.7}
\end{equation}
$\omega_{j}$ is arbitrary and in the general case has nothing
to do with the Hamiltonian.

In the Liouville picture,
\begin{equation}
A(t)=A_{L}(\gamma,t)\equiv A(\gamma,t).
\label{1.3.8}
\end{equation}
For the sake of brevity, we denote the canonical dynamical
variables by
\begin{equation}
\alpha_{j}(\gamma)\equiv A(\gamma) \quad  {\rm where}\quad
 A(\gamma)=
\alpha_{j}, \qquad \alpha_{j}^{*}(\gamma)=A^{*}(\gamma);
\label{1.3.9}
\end{equation}
$\alpha_{j}(\gamma),\:\alpha_{j}^{*}(\gamma)$ are time
independent. We have\begin{equation}
[\alpha_{j},\alpha_{k}]=0,\;[\alpha_{j}^{*},
\alpha_{k}^{*}]=0,\;[\alpha_{j},\alpha_{k}^{*}]=\delta_{jk}.
\label{1.3.10}
\end{equation}

In the Hamilton picture,
\begin{equation}
A_{t}(t)=A_{Ht}(\gamma,t)\equiv A_{t}(\gamma,t)=
A(\gamma_{t},t).
\label{1.3.11}
\end{equation}
The equations
\begin{equation}
\frac{d\alpha_{jt}}{dt}=-{\rm i}\frac{\partial
H(\gamma_{t},t)}{\partial \alpha_{jt}^{*}},\qquad
\frac{d\alpha_{jt}^{*}}{dt}={\rm i}
\frac{\partial H(\gamma_{t},t)}{\partial \alpha_{jt}}
\label{1.3.12}
\end{equation}
generate
\begin{equation}
\gamma_{t}={\cal G}_{t,t_{0}}\gamma;
\label{1.3.13}
\end{equation}
${\cal G}_{t,t_{0}}\gamma$ gives for a fixed $\gamma$ a curve
and for a fixed $t$ a transformation $\Gamma\to\Gamma$;
${\cal G}_{t,t_{0}}^{-1}$ is the inverse transformation.

We have
\begin{equation}
\int d\gamma\rho_{t}(\gamma)A(\gamma,t)=
\int d\gamma\rho(\gamma)A_{t}(\gamma,t),
\label{1.3.14}
\end{equation}
\begin{equation}
\rho(\gamma)\equiv \rho_{H}(\gamma),\qquad
\rho_{t}(\gamma)\equiv \rho_{Lt}(\gamma)=
\rho({\cal G}_{t,t_{0}}^{-1}\gamma),
\label{1.3.15}
\end{equation}
\begin{equation}
A_{t}(\gamma,t)=
A({\cal G}_{t,t_{0}}\gamma,t).
\label{1.3.16}
\end{equation}

\subsection{A quantum system}

A separable Hilbert space ${\cal H}$ is given. A dynamical
variable $A$ is an operator, $\rho$ is the statistical operator,
\begin{equation}
\left\langle \rho,A \right\rangle={\rm Tr}\left\{
\rho A \right\}.
\label{1.4.1}
\end{equation}

In the Schr\"odinger picture,
\begin{equation}
A_{S}(t)\equiv A(t),\qquad \rho_{St}\equiv \rho_{t}.
\label{1.4.2}
\end{equation}

In the Heisenberg picture,
\begin{equation}
A_{Ht}(t)\equiv A_{t}(t),\qquad \rho_{H}\equiv\rho.
\label{1.4.3}
\end{equation}

We have
\begin{equation}
\rho_{t}=U_{t,t_{0}}\rho U_{t_{0},t},\qquad
A_{t}(t)=U_{t_{0},t}A(t)U_{t,t_{0}},
\label{1.4.4}
\end{equation}
\begin{equation}
U_{t,t_{0}}=T\exp\left\{ -{\rm i}
\int\limits_{t_{0}}^{t}H(t')dt' \right\},\qquad
H(t)\equiv H_{S}(t).
\label{1.4.5}
\end{equation}

The Hilbert space ${\cal H}$ may be realized as the Fock space
constructed on the annihilation and creation operators
$a_{j},\; a_{j}^{\dagger}$,
\begin{equation}
[a_{j},a_{k}]=0,\qquad [a_{j}^{\dagger},a_{k}^{\dagger}]=0,
\qquad [a_{j},a_{k}^{\dagger}]=\delta_{jk};
\label{1.4.6}
\end{equation}
then
\begin{equation}
A(t)=A(\nu,t),\qquad \nu=\left\{ (a_{j},a_{j}
^{\dagger}):j\in J \right\}.
\label{1.4.7}
\end{equation}

\subsection{Classical-quantum relation}

The classical-quantum relation is given by
\begin{equation}
\alpha_{j}\leftrightarrow a_{j}, \quad \alpha_{j}^{*}
\leftrightarrow a_{j}^{\dagger},
\label{1.5.1}
\end{equation}
\begin{equation}
A(\gamma,t)\leftrightarrow A(\nu,t).
\label{1.5.2}
\end{equation}

\section{Field}

\subsection{Field, momentum, and Hamiltonian}

In a comoving reference frame, metric is of the form
\begin{equation}
g=g(x,t)=(dt)^{2}-h_{ik}(x,t)dx^{i}dx^{k},
\label{2.1.1}
\end{equation}
and the Hamiltonian in the Liouville/Schr\"odinger picture is
\begin{equation}
H(t)=\frac{1}{2}\int_{S}dx\sqrt{|h(x,t)|}\left\{
\pi^{2}(x,t)+h^{ik}(x,t)\partial_{i}\phi(x,t)
\partial_{k}\phi(x,t)+m^{2}\phi^{2}(x,t) \right\}
\label{2.1.2}
\end{equation}
where the field $\phi(x,t)$ and the momentum $\pi(x,t)$ are
dynamical variables.

\subsection{The problems of phase/Hilbert space and
field-momentum}

In classical field theory, we should define a phase space
$\Gamma$ and the dynamical variables $\phi,\;\pi$ as functions
on it; in quantum field theory, we should fix a Hilbert space  $
{\cal H}$ and the operators $\phi,\;\pi$ in it.

\subsection{A straightforward approach}

In classical theory, a straightforward approach would
be as follows:
\begin{equation}
\gamma=\left\{ (\alpha_{s},\alpha_{s}^{*}):
s\in S \right\}\leftrightarrow\left\{ (\phi_{s},
\pi_{s}):s\in S \right\}\equiv\chi,
\label{2.3.1}
\end{equation}
\begin{equation}
\phi(\chi,s,t)=\phi_{s},\qquad \pi(\chi,s,t)=\pi_{s}
\label{2.3.2}
\end{equation}
(cf. (\ref{1.3.9})). But there is no canonical,
i.e., natural way for introducing manifold structure in
the set $\left\{ \chi \right\}$.

In quantum theory, there is no canonical way for choosing
the representation of $\phi,\;\pi$, the commutation relations
only being given.

\subsection{Mode approach}

In the classical case, we put
\begin{equation}
\gamma=\left\{ (\alpha_{j},\alpha_{j}^{*}):
j\in J \right\},\qquad \|J\|=\aleph_{0}\quad ({\rm or}\; \aleph),
\label{2.4.1}
\end{equation}
\begin{equation}
\Gamma=l_{2}\quad ({\rm or}\; L_{2}),
\label{2.4.2}
\end{equation}
\begin{equation}
\phi(\gamma,x,t)=\frac{1}{\sqrt{2}}\sum_{j}
\left\{ \frac{1}{\sqrt{\omega_{j}(t)}}u_{j}(x,t)
\alpha_{j}+\frac{1}{\sqrt{\omega_{j}^{*}(t)}}
u_{j}^{*}(x,t)\alpha_{j}^{*} \right\}\quad ({\rm or}\;\int dj),
\label{2.4.3}
\end{equation}
\begin{equation}
\pi(\gamma,x,t)=\frac{{\rm i}}{\sqrt{2}}\sqrt{\frac{
|h_{u}(x,t)|}{|h(x,t)|}}\sum_{j}\left\{ -\sqrt{\omega_{j}^{*}}
u_{j}(x,t)\alpha_{j}+\sqrt{\omega_{j}(t)}
u_{j}^{*}(x,t)\alpha_{j}^{*} \right\},
\label{2.4.4}
\end{equation}
where
\begin{equation}
u_{j}^{*}=u_{p(j)},\quad \omega_{p(j)}=\omega_{j},
\label{2.4.5}
\end{equation}
$p$ is a permutation, such that
\begin{equation}
p\circ p={\rm I},\quad p^{-1}=p,
\label{2.4.6}
\end{equation}
and
\begin{equation}
\int_{S}dx\sqrt{|h_{u}(x,t)|}u_{j}^{*}(x,t)u_{k}(x,t)=
\delta_{jk},
\label{2.4.7}
\end{equation}
so that we obtain the canonical commutation relations:
\begin{equation}
\begin{array}{l}
[\phi_{t}(x_{1},t),\phi_{t}(x_{2},t)]=0,
\qquad [\pi_{t}(x_{1},t),\pi_{t}(x_{2},2)]=0,\\
\quad [\phi_{t}(x_{1},t),\pi_{t}(x_{2},t)]=
{\rm i}\sqrt{\frac{|h_{u}(x_{2},t)|}
{|h(x_{2},t)|}}\sum_{j}u_{j}(x_{1},t)u_{j}^{*}(x_{2},t)=
{\rm i}\frac{\delta(x_{1}-x_{2})}
{\sqrt{|h(x_{2},t)|}}={\rm i}\delta_{h(s_{2},t)}(s_{1},s_{2}),\\
\qquad  s_{1},s_{2}\in S.
\end{array}
\label{2.4.8}
\end{equation}

In the quantum case, we substitute $\nu$ for $\gamma$:
\begin{equation}
\gamma\to \nu.
\label{2.4.9}
\end{equation}

\subsection{The problem of basic elements}

The basic elements of the construction being developed are
modes $u_{j}^{}$'s and ``frequencies'' $\omega_{j}^{}$'s.
The problem is in choosing a family of them,
\begin{equation}
\{(u_{j},\omega_{j}):j\in J\}.
\label{2.5.1}
\end{equation}

\section{The conventional construction}

\subsection{Basic elements, field and momentum}

The conventional construction is realized by the following
choice of the basic elements:
\begin{equation}
\frac{\partial u_{j}}{\partial t}=0,\quad \frac{\partial
\omega_{j}}{\partial t}=0,\quad \frac{\partial h_{u}}
{\partial t}=0,
\label{3.1.1}
\end{equation}
\begin{equation}
\int_{S}dx \sqrt{|h_{u}(x)|}u_{j}^{*}(x)u_{k}(x)=\delta_{jk},
\label{3.1.2}
\end{equation}
otherwise the choice is arbitrary.

Thus in the classical case
\begin{equation}
\phi(\gamma,x,t)=\phi(\gamma,x)=\frac{1}{\sqrt{2}}\sum_{j}
\left\{ \frac{1}{\sqrt{\omega_{j}}}u_{j}(x)\alpha_{j}
+\frac{1}{\sqrt{\omega_{j}^{*}}}u_{j}^{*}(x)\alpha_{j}^{*}
 \right\},
\label{3.1.3}
\end{equation}
\begin{equation}
\pi(\gamma,x,t)=\pi(\gamma,x)=\frac{{\rm i}}{\sqrt{2}}
\sqrt{\frac{|h_{u}(x)|}{|h(x,t)|}}\sum_{j}
\left\{ -\sqrt{\omega_{j}^{*}}u_{j}(x)\alpha_{j}+
\sqrt{\omega_{j}}u_{j}^{*}(x)\alpha_{j}^{*} \right\}.
\label{3.1.4}
\end{equation}

The quantum case is obtained by the substitution (\ref{2.4.9}).

\subsection{Dynamics}

We have
\begin{equation}
\frac{\partial \phi}{\partial t}=0,\quad
\frac{\partial (\sqrt{h}\pi)}{\partial t}=0,
\label{3.2.1}
\end{equation}
which implies that the Klein-Gordon equation is fulfilled
in the Hamilton/Heisenberg picture,
\begin{equation}
(\Box +m^{2})\phi_{t}=0.
\label{3.2.2}
\end{equation}

Furthermore,
\begin{equation}
\alpha_{jt}=\sum_{l}[\xi_{jlt}\alpha_{l}+\eta_{jlt}\alpha_{l}
^{*} ],
\label{3.2.3}
\end{equation}
so that
\begin{equation}
\phi_{t}(\gamma,x)=\sum_{l}\left\{ f_{lt}\alpha_{l}
+f_{lt}^{*}\alpha_{l}^{*} \right\},\qquad  {\rm class.},
\label{3.2.4}
\end{equation}
\begin{equation}
\phi_{t}(\nu,x)=\sum_{l}\left\{ f_{lt}a_{l}
+f_{lt}^{*}a_{l}^{*} \right\},\qquad {\rm quant.},
\label{3.2.5}
\end{equation}
the functions $f_{lt}^{}$'s meeting the equations
\begin{equation}
(\Box+m^{2})f_{lt}=0,
\label{3.2.6}
\end{equation}
\begin{equation}
(f_{lt},f_{l't})_{\Omega}=\delta_{ll'}\qquad
{\rm and\; so\; on},
\label{3.2.7}
\end{equation}
where $(\cdot,\cdot)_{\Omega}$ is the symplectic inner product
of solutions to the Klein-Gordon equation [1,2,5].

\subsection{Particle creation}

Let us put
\begin{equation}
h_{u}=h(t_{0}),
\label{3.3.1}
\end{equation}
\begin{equation}
\triangle(t_{o})u_{j}=-k_{j}^{2}u_{j},
\label{3.3.2}
\end{equation}
\begin{equation}
\omega_{j}=\sqrt{k_{j}^{2}+m^{2}}.
\label{3.3.3}
\end{equation}
Then in the quantum case, using the normal ordering we obtain
\begin{equation}
H(t_{0})=\sum_{j}\omega_{j}a_{j}^{\dagger}a_{j}.
\label{3.3.4}
\end{equation}
Let
\begin{equation}
\rho=\rho_{t_{0}}=|{\rm vac\rangle\langle
vac}|,\quad a_{j}
|{\rm vac}\rangle=0.
\label{3.3.5}
\end{equation}
We have
\begin{equation}
H_{t_{0}}(t_{0})|{\rm vac\rangle}=H(t_{0})|{\rm vac}\rangle=0.
\label{3.3.6}
\end{equation}
But we obtain for $t\ne t_{0}$
\begin{equation}
H_{t}(t)|{\rm vac}\rangle\ne 0,
\label{3.3.7}
\end{equation}
which may be interpreted as particle creation,
the result being independent of measurements.

\subsection{Ambiguity: Uncountable set of
nonequivalent field systems}

In view of subsection 3.1, the conventional construction
involves an uncountable set of field systems which are
nonequivalent with respect to the Bogolubov transformations:
The families $\left\{ f_{lt} \right\}$ relating to different
systems are not connected by those transformations. To see this,
it suffices to change one frequency in eq.(\ref{3.1.3}). It
should be particularly emphasized that this ambiguity has
nothing to do with different representations of the same field
$\phi_{t}(x)$ using the Bogolubov transformations or with
canonical/unitary transformations. Even restricting ourselves
to the choice given by eqs.(\ref{3.3.1})-(\ref{3.3.3}),
in view of the arbitrariness of $t_{0}$, we do not eliminate
the ambiguity.

Thus the canonical construction does not result in a definite
field theory.

\section{The canonical theory}

The central idea of the canonical theory is as follows.
The choice of the basic elements should be canonical, or
natural: It should involve only spacetime structure, which
is given by spacetime manifold topology and metric.

\subsection{Product space-time}

The employment of the comoving reference frame implies that
spacetime manifold $M$ is a trivial bundle [6], so that we
assume from the outset that $M$ is the trivial bundle, i.e.,
product space-time,
\begin{equation}
M=T\times S,\quad M\ni p=(t,s),\quad t\in T,\quad s\in S,
\label{4.1.1}
\end{equation}
where $T$ is the cosmic time and $S$ is the cosmic 3-space.

Metric in the comoving reference frame is of the form
\begin{equation}
g=g(s,t)=dt\otimes dt- h(t)=(dt)^{2}-
h_{ik}(x,t)dx^{i}dx^{k}.
\label{4.1.2}
\end{equation}

\subsection{Basic elements, field, momentum, and Hamiltonian}

In the Liouville/Schr\"odinger picture, the Hamiltonian may be
presented as
\begin{equation}
H(t)=\frac{1}{2}\int_{S}dx\sqrt{|h(x,t)|}\left\{
\pi^{2}(x,t)-\phi(x,t)\triangle \phi(x,t)+m^{2}
\phi^{2}(x,t) \right\}.
\label{4.2.1}
\end{equation}
A natural, or canonical choice of the basic elements is as
follows. We put
\begin{equation}
h_{u}=h,
\label{4.2.2}
\end{equation}
\begin{equation}
\triangle(t)u_{j}=-k_{j}^{2}(t)u_{j},\quad u_{j}=u_{j}
(x,t),
\label{4.2.3}
\end{equation}
\begin{equation}
\omega_{j}=\omega_{j}(t)=\sqrt{k_{j}^{2}(t)+m^{2}}.
\label{4.2.4}
\end{equation}
Now the scalar product is defined by
\begin{equation}
(\varphi_{1},\varphi_{2})_{t}=\int_{S}
dx\sqrt{|h(x,t)|}\varphi_{1}^{*}(x)\varphi_{2}(x),
\label{4.2.5}
\end{equation}
and
\begin{equation}
(u_{j},u_{k})_{t}=\delta_{jk}.
\label{4.2.6}
\end{equation}

We obtain in the classical and quantum cases
\begin{equation}
\phi(\gamma,s,t)=\frac{1}{\sqrt{2}}\sum_{j}
\frac{1}{\sqrt{\omega_{j}(t)}}\left\{
u_{j}(s,t)\alpha_{j}+u_{j}^{*}(s,t)\alpha_{j}^{*} \right\},
\label{t4.2.7}
\end{equation}
\begin{equation}
\pi(\gamma,s,t)=\frac{{\rm i}}{2} \sum_{j}\sqrt{\omega_{j}
(t)}\left\{ -u_{j}(s,t)\alpha_{j}+
u_{j}^{*}(s,t)\alpha_{j}^{*} \right\},
\label{4.2.8}
\end{equation}
\begin{equation}
H(\gamma,t)=\sum_{j}\omega_{j}(t)\alpha_{j}^{*}\alpha_{j},
\label{4.2.9}
\end{equation}
and
\begin{equation}
\phi(\nu,s,t)=\frac{1}{\sqrt{2}}\sum_{j}
\frac{1}{\sqrt{\omega_{j}(t)}}\left\{
u_{j}(s,t)a_{j}+u_{j}^{*}(s,t)a_{j}^{\dagger} \right\},
\label{t4.2.10}
\end{equation}
\begin{equation}
\pi(\nu,s,t)=\frac{{\rm i}}{2} \sum_{j}\sqrt{\omega_{j}
(t)}\left\{ -u_{j}(s,t)a_{j}+
u_{j}^{*}(s,t)a_{j}^{\dagger} \right\},
\label{4.2.11}
\end{equation}
\begin{equation}
H(\nu,t)=\sum_{j}\omega_{j}(t)a_{j}^{\dagger}a_{j}
\qquad {\rm (normal\;ordering)},
\label{4.2.12}
\end{equation}
respectively.

\subsection{Dynamics}

In the classical case, dynamics is determined by the
dynamical variables $\alpha_{jt},\;\alpha_{jt}^{*}$.
We obtain from eqs.(\ref{1.3.12}),(\ref{4.2.9})
\begin{equation}
\alpha_{jt}={\rm e}^{-{\rm i}\beta_{j}(t,t_{0})}\alpha_{j},
\quad \alpha_{jt}^{*}={\rm e}^{{\rm i}\beta_{j}(t,t_{0})}
\alpha_{j}^{*},
\label{4.3.1}
\end{equation}
\begin{equation}
\beta_{j}(t,t_{0})=\int\limits_{t_{0}}^{t}
\omega_{j}(t')dt'.
\label{4.3.2}
\end{equation}
Thus the field in the Hamilton picture is
\begin{equation}
\begin{array}{l}
\phi_{t}(\gamma,s,t)=\frac{1}{2}\sum_{j}
\frac{1}{\sqrt{\omega_{j}(t)}}\left\{
u_{j}(s,t)\alpha_{jt}+u_{j}^{*}(s,t)
\alpha_{jt}^{*} \right\}\\
\qquad \qquad  {}=\frac{1}{2}\sum_{j}\frac{1}{\sqrt{\omega_{j}(t)}}
\left\{ u_{jt}(s,t)\alpha_{j}+
u_{jt}^{*}(s,t)\alpha_{j}^{*} \right\}
\end{array}
\label{4.3.3}
\end{equation}
where
\begin{equation}
u_{jt}(s,t)={\rm e}^{-{\rm i}\beta_{j}(t,t_{0})}u_{j}(s,t).
\label{4.3.4}
\end{equation}
The Hamiltonian
\begin{equation}
H_{t}(\gamma,t)=H(\gamma,t)=\sum_{j}\omega_{j}(t)
\alpha_{j}^{*}\alpha_{j}.
\label{4.3.5}
\end{equation}

In the quantum case, with regard to
\begin{equation}
[H(t_{1}),H(t_{2})]=0,
\label{4.3.6}
\end{equation}
we obtain for the time evolution operator (\ref{1.4.5})
\begin{equation}
U(t,t_{0})=\exp\{-{\rm i}\int\limits_{t_{0}}^{t}
H(t')dt'\}=\prod _{j}{\rm e}^{-{\rm i}
\beta_{j}(t,t_{0})a_{j}^{\dagger}a_{j}}.
\label{4.3.7}
\end{equation}
We find in the Heisenberg picture
\begin{equation}
a_{jt}={\rm e}^{-{\rm i}\beta_{j}(t,t_{0})}a_{j},
\quad a_{jt}^{\dagger}={\rm e}^{{\rm i}\beta_{j}(t,t_{0})}
a_{j}^{\dagger},
\label{4.3.8}
\end{equation}
and the field operator
\begin{equation}
\phi_{t}(\nu,s,t)=\frac{1}{\sqrt{2}}\sum_{j}
\frac{1}{\sqrt{\omega_{j}(t)}}\left\{
u_{jt}(s,t)a_{j}+u_{jt}^{*}(s,t)a_{j}^{\dagger} \right\}.
\label{4.3.9}
\end{equation}
The Hamiltonian
\begin{equation}
H_{t}(\nu,t)=H(\nu,t)=\sum_{j}\omega_{j}(t)a_{j}^{\dagger}
a_{j}.
\label{4.3.10}
\end{equation}

\subsection{Particles}

We have
\begin{equation}
H(t)=\sum_{j}\omega_{j}(t)N_{j},
\label{4.4.1}
\end{equation}
where in the quantum case
\begin{equation}
N_{j}=a_{j}^{\dagger}a_{j}
\label{4.4.2}
\end{equation}
is the occupation number operator. Thus the canonical theory
implies the existence of particles,
\begin{equation}
\omega_{j}(t)=\sqrt{k_{j}^{2}(t)+m^{2}}
\label{4.4.3}
\end{equation}
being a particle energy.

\subsection{No particle creation}

We have
\begin{equation}
N_{jt}=N_{j},\qquad \frac{dN_{jt}}{dt}=0.
\label{4.5.1}
\end{equation}
Thus there is no particle creation.

\subsection{The violation of the Klein-Gordon equation}

With eqs.(\ref{4.3.3}),(\ref{4.3.9}),(\ref{4.3.4})
in mind, we find
\begin{equation}
(\Box+m^{2})\left[ \frac{u_{jt}}{\sqrt{\omega_{j}}}\right]
=\frac{1}{\sqrt{|h|}}\frac{\partial \sqrt{|h|}}{\partial t}
\frac{\partial}{\partial t}\left[ \frac{u_{jt}}
{\sqrt{\omega_{j}}} \right]+\frac{\partial}{\partial t}
\left\{ \frac{\partial}{\partial t}\left[ \frac{u_{j}}
{\sqrt{\omega_{j}}} \right]{\rm e}^
{-{\rm i}\beta_{j}} \right\}-
{\rm i}\frac{\partial}{\partial t}[\sqrt{\omega_{j}}u_{j}]
{\rm e}^{-{\rm i}\beta_{j}}.
\label{4.6.1}
\end{equation}
Thus the Klein-Gordon equation is violated in the generic case
of a nonstationary metric.

The Klein-Gordon equation being abandoned, the equations of
motion are those in the Liouville/Schr\"odinger and
Hamilton/Heisenberg pictures.

\subsection{Relativistic-gravitational nonlocality}

A local change in the metric $h$ results in changing the Laplacian
$\triangle$ and, by the same token, solutions to the equation
(\ref{4.2.3}), i.e., $k_{j}^{2},\;u_{j},\;\omega_{j}$, and
$u_{jt}/\sqrt{\omega_{j}}$. We call this phenomenon
relativistic-gravitational nonlocality.

Generally, relativistic-gravitational nonlocality means that
\begin{equation}
A(\gamma/\nu,s,t)=A[\gamma/\nu,s;h(t)],
\label{4.7.1}
\end{equation}
i.e., that a Liouville/Schr\"odinger dynamical variable
at a point $p=(s,t)$ depends on
the metric $h(t)$ in the whole 3-space $S$.

The degree of quantum-gravitational nonlocality may be
characterized
by the quantity
\begin{equation}
b=\frac{\partial }{\partial t}\left.\left[ \frac{u}{\sqrt
{\omega}}\right]\right/
\left[ \frac{u}{\sqrt{\omega}}
\right]\omega=\left.\frac{\partial u}{\partial t}\right/u\omega
-\left.\frac{1}{2}\frac{d\omega}{dt}\right/\omega^{2}.
\label{4.7.2}
\end{equation}
We have
\begin{equation}
\frac{d\omega}{dt}=\frac{k}{\omega}\frac{dk}{dt},
\label{4.7.3}
\end{equation}
so that
\begin{equation}
b=\left.\frac{\partial u}{\partial t}\right/u\omega-
\frac{k}{2\omega^{3}}\frac{dk}{dt}.
\label{4.7.4}
\end{equation}

\subsection{The geometric
principle as an extension of the principle of covariance }

Nonlocality is incompatible with the local principle of
covariance. More general than the latter is the geometric
principle: Spacetime structure and dynamical equations should
be phrased in a geometric, coordinate-independent form.
The principle of covariance is a
local version of the geometric principle.

The canonical theory meets the geometric principle.

\subsection{The energy-momentum tensor}

For the sake of brevity, from this point on we consider
the quantum field. The corresponding results for the classical
field are obtained in an obvious way.

Normal ordering on the energy-momentum tensor in the
comoving reference frame produces
\begin{equation}
T_{00}=\frac{1}{2}:\left\{ \pi^{2}+h^{ik}
\partial_{i}\phi\partial_{k}\phi+m^{2}\phi^{2} \right\}:\;,
\label{4.9.1}
\end{equation}
\begin{equation}
H(t)=\int_{S}dx\sqrt{|h(t)|}T_{00},
\label{4.9.2}
\end{equation}
\begin{equation}
\begin{array}{l}
T_{ik}=:\left\{\partial_{i}\phi\partial_{k}\phi+
\frac{1}{2}h_{ik}[\pi^{2}-h^{lm}\partial_{l}\phi
\partial_{m}\phi-m^{2}\phi^{2}]  \right\}:\\
\qquad {}=:\partial_{i}\phi\partial_{k}\phi:+
h_{ik}[:\pi^{2}:-T_{00}],
\end{array}
\label{4.9.3}
\end{equation}
and for a mean value
\begin{equation}
(\Psi,T_{ik}\Psi)=(\Psi,:\partial_{i}\phi\partial_{k}\phi:\Psi)
+h_{ik}(\Psi,[:\pi^{2}:-T_{00}]\Psi).
\label{4.9.4}
\end{equation}

\section{Applications to cosmology}

\subsection{The metric-consistent energy-momentum tensor}

Let in eq.(\ref{4.9.4})
\begin{equation}
(\Psi,:\partial_{i}\phi\partial_{k}\phi:\Psi)\propto h_{ik}
\label{5.1.1}
\end{equation}
hold, i.e.,
\begin{equation}
(\Psi,:\partial_{i}\phi\partial_{k}\phi:\Psi)=Ch_{ik}h^{lm}
(\Psi,:\partial_{l}\phi\partial_{m}\phi:\Psi).
\label{5.1.2}
\end{equation}
Since
\begin{equation}
h^{ik}h_{ik}=3,
\label{5.1.3}
\end{equation}
we find
\begin{equation}
C=\frac{1}{3}
\label{5.1.4}
\end{equation}
and by eqs.(\ref{4.9.4}),(\ref{4.9.1})
\begin{equation}
(\Psi,T_{ik}\Psi)=\frac{1}{3}h_{ik}
(\Psi,\{ 2:\pi^{2}:-T_{00}-m^{2}:\phi^{2}:\}\Psi).
\label{5.1.5}
\end{equation}

\subsection{A homogeneous state}

Let $\Psi$ be a homogeneous state, so that
\begin{equation}
\begin{array}{l}
(\Psi,\{2:\pi^{2}:-T_{00}-m^{2}:\phi^{2}:\}\Psi)=
\frac{1}{V}\int_{S}dx\sqrt{|h|}(\Psi,\{2:\pi^{2}:-
T_{00}-m^{2}:\phi^{2}:\}\Psi),\\
\qquad {}V=V_{t}=\int_{S}dx\sqrt{|h(t)|}.
\end{array}
\label{5.2.1}
\end{equation}
We have
\begin{equation}
\int_{S}dx\sqrt{|h|}T_{00}=\sum_{j}\omega_{j}N_{j},
\label{5.2.2}
\end{equation}
\begin{equation}
\int_{S}dx\sqrt{|h|}:\pi^{2}:=\sum_{j}\omega_{j}N_{j}+
\{aa+a^{\dag}a^{\dag}\},
\label{5.2.3}
\end{equation}
\begin{equation}
\int_{S}dx\sqrt{|h|}:\phi^{2}:=\sum_{j}\frac{1}{\omega_{j}}N_{j}
+\{aa+a^{\dag}a^{\dag}\}.
\label{5.2.4}
\end{equation}

Let
\begin{equation}
N_{j}\Psi=n_{j}\Psi\qquad {\rm for\;all}\;j,
\label{5.2.5}
\end{equation}
then
\begin{equation}
(\Psi,T_{ik}\Psi)=h_{ik}\frac{1}{3V}\sum_{j}\frac{\omega_
{j}^{2}-m^{2}}{\omega_{j}}n_{j}.
\label{5.2.6}
\end{equation}
Thus the pressure is
\begin{equation}
p=\frac{1}{3V}\sum_{j}\frac{\omega_{j}^{2}-m^{2}}{\omega_{j}}n_{j}
=\frac{1}{3V}\sum_{j}\frac{k_{j}^{2}}{\omega_{j}}n_{j},
\label{5.2.7}
\end{equation}
whereas the energy density is
\begin{equation}
\rho=\frac{E}{V}=\frac{1}{V}\sum_{j}\omega_{j}n_{j}.
\label{5.2.8}
\end{equation}

\subsection{The Robertson-Walker spacetime}

For the Robertson-Walker spacetime, the metric is of the form
\begin{equation}
h(s,t)=R^{2}(t)\kappa(s),\qquad {\rm or}\quad h_{ik}=
R^{2}(t)\kappa_{ik},
\label{5.3.1}
\end{equation}
so that we have
\begin{equation}
|h|=|\kappa|R^{6},\quad |\kappa|={\rm det}(\kappa_{ik}),
\quad \sqrt{|h|}=R^{3}\sqrt{|\kappa|},\quad h^{ik}=
\frac{\kappa^{ik}}{R^{2}},
\label{5.3.2}
\end{equation}
and
\begin{equation}
\triangle=\frac{1}{R^{2}}\triangle_{\kappa},\qquad
\triangle_{\kappa}\varphi=\frac{1}{\sqrt{|\kappa|}}\partial_{i}
\left[ \sqrt{|\kappa|}\kappa^{ik}\partial_{k}\varphi \right].
\label{5.3.3}
\end{equation}
The equation (\ref{4.2.3}) results in
\begin{equation}
\frac{1}{R^{2}(t)}\triangle_{\kappa}u_{j}=-k_{j}^{2}u_{j},
\label{5.3.4}
\end{equation}
so that, in view of eq.(\ref{4.2.6}),
\begin{equation}
\triangle_{\kappa}u_{j}=-\mu_{j}^{2}u_{j},\quad
\mu_{j}^{2}={\rm const},\quad k_{j}^{2}(t)=
\frac{\mu_{j}^{2}}{R^{2}(t)},\quad u_{j}(s,t)=
\frac{1}{R^{3/2}(t)}u_{j}^{0}(s),
\label{5.3.5}
\end{equation}
and
\begin{equation}
\omega_{j}=\left[ m^{2}+\frac{\mu_{j}^{2}}{R^{2}(t)}
 \right]^{1/2},
\label{5.3.6}
\end{equation}
the last relation being a familiar result of cosmology.

In eq.(\ref{4.7.4}) we obtain
\begin{equation}
u=\frac{u^{0}(s)}{R^{3/2}(t)},\quad k=\frac{\mu}{R(t)},
\label{5.3.7}
\end{equation}
so that
\begin{equation}
|b|=\frac{3}{2\omega}\frac{dR/dt}{R}
-\frac{1}{2}\frac{(\mu/R)^{2}}{\omega^{3}}
\frac{dR/dt}{R}=\frac{3H}{2\omega}-
\frac{1}{2}\frac{k^{2}}{\omega^{3}}H=
\left(3-\frac{k^{2}}{\omega^{2}}\right)\frac{H}{2\omega}
<\frac{3H}{2\omega},
\label{5.3.8}
\end{equation}
where $H$ is the Hubble constant.
\begin{equation}
{\rm For}\quad H\approx\frac{1}{3}10^{-17}c^{-1}
\quad {\rm and}\quad  \omega\sim 10^{15}c^{-1},\qquad |b|<10^{-32}.
\label{5.3.9}
\end{equation}

With eqs.(\ref{5.2.7}),(\ref{5.2.8}) in mind, we have
\begin{equation}
k_{j}^{2}=\frac{\nu_{j}^{2}}{V^{2/3}},\quad
\nu_{j}^{2}={\rm const},\qquad \omega_{j}=\left(
m^{2}+\frac{\nu_{j}^{2}}{V^{2/3}}\right)^{1/2},
\label{5.3.10}
\end{equation}
so that we find
\begin{equation}
\frac{dE}{dV}=\frac{d(\rho V)}{dV}=\sum_{j}n_{j}
\frac{d\omega_{j}}{dV}=-\frac{1}{3V}\sum_{j}n_{j}
\frac{k_{j}^{2}}{\omega_{j}}=-p,
\label{5.3.11}
\end{equation}
i.e.,
\begin{equation}
dE=-pdV,
\label{5.3.12}
\end{equation}
which is a standard relation.

\subsection{Universe dynamics}

In this and the next subsections, we follow the papers [7,8].

The $S$-projected Einstein equation yields
\begin{equation}
G_{ik}=8\pi\kappa_{g}(\Psi,T_{ik}\Psi)\quad \Rightarrow\quad
2\ddot R R+\dot R^{2}+1=-8\pi\kappa_{g} pR^{2}
\label{5.4.1}
\end{equation}
where $\kappa_{g}$ is the gravitational constant; eq.(\ref{5.3.12})
amounts to
\begin{equation}
\frac{d(\rho R^{3})}{dR}=-3pR^{2}.
\label{5.4.2}
\end{equation}
We obtain from eqs.(\ref{5.4.1}),(\ref{5.4.2})
\begin{equation}
\frac{d}{dR}\left(R\dot R^{2}+R-
\frac{8\pi\kappa_{g}}{3}\rho R^{3}    \right)=0,
\label{5.4.3}
\end{equation}
whence
\begin{equation}
R\dot R^{2}+R-\frac{8\pi\kappa_{g}}{3}\rho R^{3}=L=
{\rm const}.
\label{5.4.4}
\end{equation}

The length $L$, which is an integral of motion, is called
cosmic length. In accordance with this, the model considered
is called the cosmic length universe.

The Friedmann universe corresponds to a particular value of
the cosmic length,
\begin{equation}
L_{{\rm Friedmann}}=0.
\label{5.4.5}
\end{equation}
In this sense, the Friedmann universe is the zero-length universe.

The value $L=0$ results from the equation
\begin{equation}
G_{0\mu}=8\pi\kappa_{g}(\Psi,T_{0\mu}\Psi),
\label{5.4.6}
\end{equation}
which is violated by quantum jumps inherent in the generic
case of interacting quantum fields.

\subsection{Lifting the problem of the missing dark matter}

The most important problem facing modern cosmology is that of
the missing dark matter [9]. Most of the mass of galaxies and
an even larger fraction of the mass of clusters of galaxies is
dark. The problem is that even more dark matter is required to
account for the rate of expansion of the universe.

More specifically, for the Friedmann universe, the equation
\begin{equation}
\Omega_{0}=2q_{0}
\label{5.5.1}
\end{equation}
holds, where $\Omega$ is the density parameter,
\begin{equation}
\Omega=\frac{\rho}{\rho_{c}},
\label{5.5.2}
\end{equation}
$\rho_{c}$ is the critical value of $\rho$, $q$ is the
deceleration parameter,
\begin{equation}
q=-\frac{\ddot R R}{\dot R^{2}},
\label{5.5.3}
\end{equation}
and subscript 0 indicates present-day values. In particular,
if $q_{0}>1/2$, the universe is closed and $\rho_{0}>\rho_{c}$.
But observational data give $\Omega_{0}<2q_{0}$. Eq.(\ref{5.5.1})
reduces to
\begin{equation}
\Omega_{0}=1+\frac{1}{R_{0}^{2}H_{0}^{2}}.
\label{5.5.4}
\end{equation}

{}From eq.(\ref{5.4.4}) we obtain
\begin{equation}
\Omega_{0}=1+\frac{1-L/R_{0}}{R_{0}^{2}H_{0}^{2}}
\label{5.5.5}
\end{equation}
in place of eq.(\ref{5.5.4}). For
\begin{equation}
p_{0}\ll\frac{1}{3}\rho_{0},
\label{5.5.6}
\end{equation}
which is fulfilled, eq.(\ref{5.5.5}) reduces to
\begin{equation}
\Omega_{0}=2q_{0}-\frac{L/R_{0}}{R_{0}^{2}H_{0}^{2}}
\label{5.5.7}
\end{equation}
in place of eq.(\ref{5.5.1}).

Eq.(\ref{5.5.7}) lifts the problem.

\section{An application to black holes}

\subsection{The Lema\^\i tre metric}

In the case of a black hole, the metric in the comoving
reference frame is the Lema\^\i tre metric:
\begin{equation}
h=\frac{1}{[(3/2r_{s})(R-t)]^{2/3}}dR^{2}+r_{s}^{2/3}
\left[ \frac{3}{2}(R-t) \right]^{4/3}(d\theta^{2}+
\sin^{2}\theta d\varphi^{2}),
\label{6.1.1}
\end{equation}
where $r_{s}$ is the Schwarzschild radius. The Schwarzschild
coordinate is
\begin{equation}
r=\left(\frac{3}{2} \right)^{2/3}r_{s}^{1/3}(R-t)^{2/3}.
\label{6.1.2}
\end{equation}

\subsection{Quantum field in the comoving reference frame}

With the equation (\ref{4.2.3}) in mind, we find
\begin{equation}
\Delta\chi\equiv\Delta_{\vec R}\chi=\frac{1}{r^{2}}
\partial_{r}[r^{2}\partial_{r}\chi]+\frac{1}{r^{2}\sin\theta}
\partial_{\theta}[\sin\theta\partial_{\theta}\chi]+
\frac{1}{r^{2}\sin^{2}\theta}\partial_{\varphi}^{2}\chi
=\Delta_{\vec r}\chi
\label{6.2.1}
\end{equation}
where
\begin{equation}
\vec R=(R,\theta,\varphi),\qquad \vec r=(r,\theta,\varphi).
\label{6.2.2}
\end{equation}
Thus eq.(\ref{4.2.3}) reduces to
\begin{equation}
\Delta_{\vec r}u_{j}=-k_{j}^{2}u_{j},
\label{6.2.3}
\end{equation}
whence
\begin{equation}
u_{j}=u_{j}(r,\theta,\varphi)
\label{6.2.4}
\end{equation}
with $r$ given by eq.(\ref{6.1.2}), and
\begin{equation}
\frac{dk_{j}^{2}}{dt}=0,\quad \omega_{j}=\left[ m^{2}+
k_{j}^{2} \right]^{1/2},\quad \frac{d\omega_{j}}{dt}=0.
\label{6.2.5}
\end{equation}
So in the comoving reference frame
\begin{equation}
\omega_{j}={\rm const},\qquad H=\sum_{j}\omega_{j}
a_{j}^{\dag}a_{j},\quad \frac{dH}{dt}=0.
\label{6.2.6}
\end{equation}

In eq.(\ref{4.7.2}) we have
\begin{equation}
\frac{d\omega}{dt}=0,
\label{6.2.7}
\end{equation}
so that
\begin{equation}
b=\left.\frac{\partial u}{\partial t}\right/u\omega.
\label{6.2.8}
\end{equation}
We find from eq.(\ref{6.1.2})
\begin{equation}
\frac{\partial u}{\partial t}=\frac{\partial u}{\partial r}
\left(\frac{r_{s}}{r}\right)^{1/2}.
\label{6.2.9}
\end{equation}
By [10], in view of $\sqrt{|h|}\sim r^{3/2}$,
\begin{equation}
\left|\frac{\partial u}{\partial r}\right|\sim
\left(k^{2}+\frac{1}{r^{2}}\right)^{1/2}|u|,
\label{6.2.10}
\end{equation}
so that
\begin{equation}
|b|\sim\left[ \frac{k^{2}+1/r^{2}}{\omega^{2}}\:
\frac{r_{s}}{r}\right]^{1/2}=\left[
\frac{\omega^{2}-m^{2}+1/r^{2}}{\omega^{2}}\:\frac
{r_{s}}{r} \right]^{1/2}.
\label{6.2.11}
\end{equation}
In particular,
\begin{equation}
{\rm for}\; r\gg\lambda=\frac{2\pi}{k},\qquad
|b|\sim\left[ \frac{\omega^{2}-m^{2}}{\omega^{2}}
\frac{r_{s}}{r} \right]^{1/2}.
\label{6.2.12}
\end{equation}

\section*{Acknowledgment}

I would like to thank Stefan V. Mashkevich for helpful
discussions.

\end{document}